# Independent characterization of the elastic and the mixing parts of hydrogel osmotic pressure

Zefan Shao, Qihan Liu*

Department of Mechanical Engineering and Materials Science, University of Pittsburgh, Pittsburgh, PA 15213, USA

* Corresponding author: E-mail address: qihan.liu@pitt.edu (Q. Liu)

## Abstract

Osmotic pressure is the driving force for the swelling of hydrogels. The hydrogel osmotic pressure can be decomposed into two parts: the mixing part due to polymer-solvent interaction and the elastic part due to polymer chain stretching. While the two components are distinguished in existing constitutive models, they have not been independently characterized in experiments. This paper reports a novel method to independently measure these two components using a fully constrained swelling test. The test allows the crosslink density to be varied at a fixed polymer content, thus varying the elastic part independently of the mixing part. Our measurement shows that the widely used Flory-Rehner model predicts the mixing part accurately for polyacrylamide hydrogel of a wide range of swelling ratios but predicts the elastic part with one order-of-magnitude error.

## 1. Introduction

Hydrogels are smart materials that change the swelling ratio in response to various environmental stimuli, such as stress, humidity, temperature, pH, and light [1, 2]. The stimuli-responsive swelling has led to various applications such as sensors [3-6], actuators [7-9], and drug delivery systems [10-12]. The swelling of a hydrogel is influenced by various microscopic mechanisms, such as the increase in the entropy of mixing, the elasticity of stretching the polymer network, the hydrophilic interaction between the polymer and water, and the charge interaction between the dissociated groups on the polymer chain and free ions. The macroscopic effect of these mechanisms in swelling can be represented as the osmotic pressure, which is an important part of hydrogel constitutive models [13-23]. Existing models decompose the osmotic pressure into two parts: the mixing part due to polymer-solvent interaction and the elastic part due to polymer chain stretching [13-23]. However, so far, it has been impossible to independently characterize the two parts in experiments. As a compromise, it is common to assume that the elastic part follows the Flory-Rehner model, which allows the mixing part to be derived from experimental measurements [24-27].

However, the assumption is dubious since the elastic part of the Flory-Rehner model results in significant deviation when predicting swelling-dependent stiffness [28-33]. Without independent experimental characterization of the two parts, the quantitative study of hydrogel constitutive models has been greatly limited.

This paper independently characterizes the mixing and elastic parts of hydrogel osmotic pressure through a fully constrained swelling test that prevents any deformation in the hydrogel relative to a stress-free swollen state. With no deformation, the mixing part is solely determined by the polymer content in the hydrogel, and the elastic part can be independently controlled by the crosslink density. When the crosslink density is sufficiently low, the elastic part becomes negligible, and then the mixing part equals the total osmotic pressure, which can be directly measured. Once the mixing part is known, the elastic part of any crosslink density can be obtained by subtracting the mixing part from the total osmotic pressure. In contrast to earlier studies that trust the elastic part of the Flory-Rehner model and correct the mixing part [24-27], we show that in polyacrylamide hydrogel, the elastic part significantly deviates from the Flory-Rehner model, but the mixing part can be well fitted by the Flory-Rehner model over a wide range of polymer content. The same testing method can be applied to various hydrogels to quantitatively study the thermodynamics of swelling.

## 2. The osmotic pressure of a polymer gel

In a solution, osmotic pressure $\Pi$ is defined as the pressure applied to the solution to stop the solvent migration across a semi-permeable membrane (Fig. 1a) [34]. Here the semi-permeable membrane allows the solvent to diffuse through but blocks the solute. In a gel, the solute of the system, the polymer chain, is crosslinked and cannot diffuse away. Consequently, there is no need for a semi-permeable membrane, and the osmotic pressure $\Pi$ is the pressure applied on the polymer network to stop swelling (Fig. 1b). Due to the crosslinking, the osmotic pressure $\Pi$ in a gel is fundamentally different from that of a solution. In a polymer solution, adding solvent moves polymer chains apart without changing the chain configuration (Fig. 1c). Consequently, elasticity is not involved. In a gel, the crosslinks prevent the polymer chains from drifting apart. Consequently, adding solvent stretches the polymer chains and increases the elastic energy (Fig. 1d).

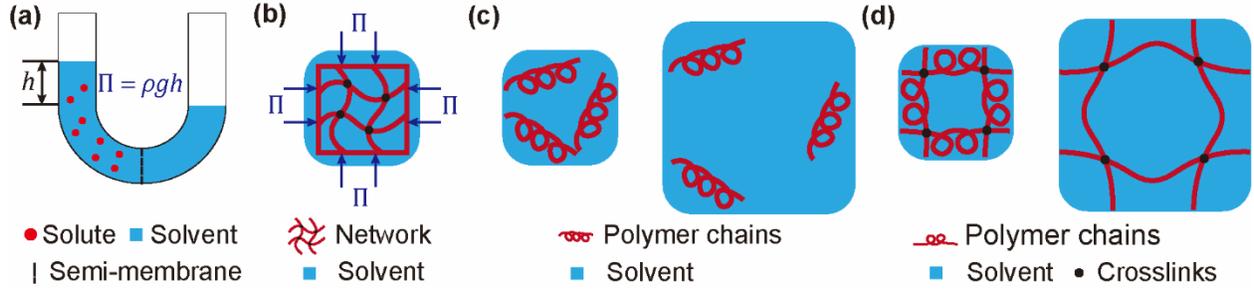

**Fig. 1 The definition of osmotic pressure. (a)** The osmotic pressure $\Pi$ in a solution is the pressure to stop the solvent from diffusing into the polymer solution across a semi-membrane. **(b)** The osmotic pressure $\Pi$ in a gel is the pressure applied on the polymer network to stop the swelling. **(c)** Mixing polymer and solvent in a solution does not involve the elasticity of the polymer chains. **(d)** Mixing polymer and solvent in a gel is penalized by the elasticity of the polymer network.

The osmotic pressure of a gel can be derived by minimizing the Helmholtz free energy $W$ per dry volume with respect to the swelling ratio $V/V_0 = 1/\phi$, where $V_0$ and $V$ are the dry volume and the swollen volume of the gel, $\phi$ is the polymer volume fraction [35]:

$$\Pi = -\frac{dW}{d(1/\phi)}. \qquad (1)$$

The free energy per dry volume $W$ is the basis of the gel constitutive models [13-22]. $W$ is divided into multiple contributions, such as chain elasticity, solvent-chain mixing, dissociation of charged groups, Donnan equilibrium of the free ions, and electromagnetic fields, depending on the physics driving the gel swelling. Here the free energy associated with chain elasticity, $W_{ela}$, and solvent-chain mixing, $W_{mix}$, are involved in the swelling of any polymer gels. Flory-Rehner model is the most widely used formulation of $W_{ela}$ and $W_{mix}$ [36]:

$$W_{ela} = \frac{1}{2}NkT[\lambda_1^2 + \lambda_2^2 + \lambda_3^2 - 3 - 2\log(\lambda_1\lambda_2\lambda_3)], \qquad (2)$$

$$W_{mix} = \frac{kT}{\Omega}\left[\frac{1-\phi}{\phi}\log(1-\phi) - \chi\phi\right]. \qquad (3)$$

Here $N$ is the number of chains per unit dry volume, $k$ is the Boltzmann constant, $T$ is the thermodynamic temperature, $\lambda_1, \lambda_2, \lambda_3$ are the principal stretches of the gel relative to the stress-free dry state, $\phi$ is the polymer volume fraction, $\Omega$ is the volume per solvent molecule, $\chi$ is a dimensionless parameter characterizing the enthalpy of mixing. Since the shear modulus of a common gel (<1MPa) is orders of magnitude lower than the bulk modulus of water (~GPa) [37, 38], a gel is practically incompressible during swelling, which leads to the constraint $\lambda_1\lambda_2\lambda_3 = 1/\phi$ [35]. If the gel is isotopically swollen, $\lambda_1 = \lambda_2 = \lambda_3 = 1/\phi^{1/3}$, then $W_{ela}$ is a function of $\phi$ alone:

$$W_{ela} = \frac{1}{2} NkT \left[ \frac{3}{\phi^{2/3}} - 3 + 2 \log(\phi) \right], \tag{4}$$

Then for a gel with $W = W_{ela} + W_{mix}$, the osmotic pressure follows from Eq. 1:

$$\Pi = \Pi_{ela} + \Pi_{mix}, \tag{5}$$

where:

$$\Pi_{ela} = -NkT \left( \phi^{\frac{1}{3}} - \phi \right), \tag{6}$$

$$\Pi_{mix} = -\frac{kT}{\Omega} [\phi + log(1 - \phi) + \chi \phi^2]. \tag{7}$$

With a constitutive model like Eqs. (6, 7), it is easy to measure different model parameters and distinguish $\Pi_{ela}$ and $\Pi_{mix}$. However, the Flory-Rehner model is known to be highly approximate and can result in large errors when compared to experiments [28-33, 39-44]. It has been proposed that the error in the Flory-Rehner model comes from $\Pi_{mix}$. Then assuming that the expression of $\Pi_{ela}$ in Eq. 6 is accurate, the correct $\Pi_{mix}$ can be obtained by $\Pi - \Pi_{ela}$[24-27]. However, whether the expression of $\Pi_{ela}$ is accurate is dubious. Consequently, it remains unclear what causes the error in the Flory-Rehner model and how to fix it.

Here we propose to independently characterize $\Pi_{mix}$ and $\Pi_{ela}$ by measuring $\Pi$ while varying the crosslink density. Since the elasticity of the polymer network originates from crosslinking, reducing the crosslink density can reduce $\Pi_{ela}$. When $|\Pi_{ela}| \ll |\Pi_{mix}|$, $\Pi = \Pi_{mix}$. Then $\Pi_{ela} = \Pi - \Pi_{mix}$ can be calculated for any other crosslinking densities. This approach allows $\Pi_{mix}$ and $\Pi_{ela}$ to be directly characterized without assuming a model.

Note that even when $|\Pi_{ela}| \ll |\Pi_{mix}|$, $\Pi_{mix}$ is different from the osmotic pressure of an uncrosslinked polymer solution. A polymer solution generally consists of many polymer chains with distributed molecular weights [45, 46]. The translational entropy of these chains can have a non-negligible effect on the osmotic pressure [47-49]. In contrast, a crosslinked polymer network is effectively a single chain with no translational entropy[50].

## 3. Measuring hydrogel osmotic pressure with constrained swelling tests

Existing studies commonly measure the hydrogel osmotic pressure with a constrained swelling test where a piece of stress-free hydrogel is constrained between two rigid plates and equilibrated with the ambient water (Fig. 2a) [51-53]. The lateral sliding of the hydrogel is constrained by the friction on the top and bottom surfaces. If the deformation near the hydrogel edge is negligible, then at thermodynamic equilibrium, the pressure measured by the load cell $p = \Pi$. However, this setup cannot directly measure $\Pi_{mix}$ because when $|\Pi_{ela}| \ll |\Pi_{mix}|$, the gel is so soft that $\Pi_{mix}$ can drive significant expansion near the hydrogel edge even if the hydrogel surface does not slide.

Here we developed a fully constrained swelling test to measure $\Pi$ when $|\Pi_{ela}| \ll |\Pi_{mix}|$ (Fig. 2b). A piece of hydrogel is constrained between a fine stainless-steel mesh and a stainless-steel die. The deformation of the hydrogel is constrained in all directions by the die and mesh. The mesh allows free water exchange. During a test, the whole setup is quickly submerged under water and preloaded to a pressure below an estimated $\Pi$. The quick submersion limits the swelling before the constraint is applied, and the preload closes any gap in the setup. The pressure on the load cell $p$ is monitored until equilibrium.

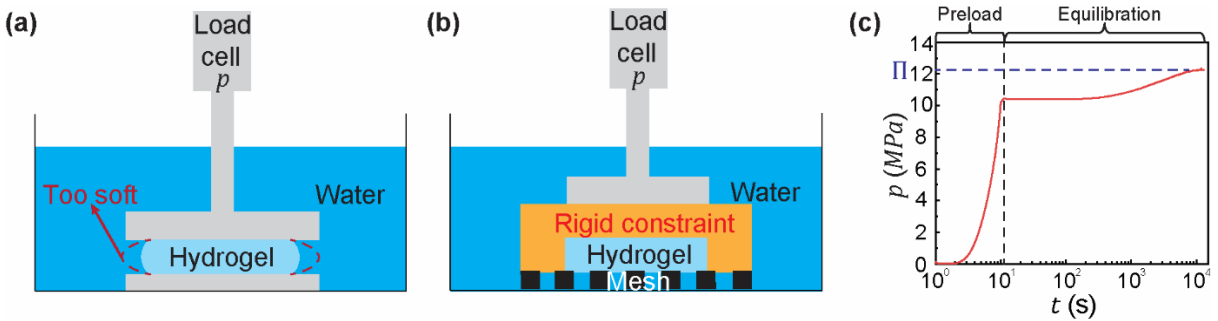

**Fig. 2 Measuring the osmotic pressure of a hydrogel. (a)** Existing studies measure osmotic pressure by swelling a piece of hydrogel between two rough plates that prevent the lateral sliding of the hydrogel. However, the method would fail if the hydrogel were so soft that the edge deforms excessively. **(b)** We measure osmotic pressure by swelling a piece of hydrogel under a fully constrained condition, which works even for extremely soft hydrogels. **(c)** A representative pressure relaxation curve is measured from a hydrogel with polymer content $\phi = 0.51$, and shear modulus $G = 60\ kPa$.

Since the equilibration time depends on the exchange of water molecules between the hydrogel and the ambient solution through the mesh, the thinner the hydrogel, the faster the water can diffuse through the hydrogel, thus the faster the equilibration. We use 1mm hydrogel disks in all our tests, which results in an equilibration time of about 3 hours. The submerging and preloading are finished in less than 30s to ensure negligible swelling before the constraint is completely applied. Fig. 2c shows a representative relaxation curve measured from a piece of polyacrylamide hydrogel with polymer content $\phi = 0.51$ and shear modulus $G = 60 kPa$ (Appendix). After equilibration, the hydrogel disks are weighted. The constraint is considered sufficient if the mass change is less than 5%.

## 4. Independently characterize the mixing and elastic parts of the osmotic pressure

Here we use the constrained swelling test to independently characterize $\Pi_{mix}$ and $\Pi_{ela}$ in polyacrylamide (pAAm) hydrogel of various polymer content and crosslink density (Appendix). Since the crosslinkers cannot be fully reacted and may not be uniformly distributed in the network, the effective crosslink density generally differs from the number of crosslinkers added in the polymer network [54, 55]. Additionally, the effective crosslink density can also be contributed by the entanglements [56-58], which cannot be predicted or directly measured. Consequently, the exact crosslink density of the hydrogel is unknown. On the other hand, the shear modulus $G$ of the hydrogel at the swollen state always increases with the crosslink density [59] and can be readily measured by a pure shear test [60]. Consequently, we will use $G$ as an indicator of the crosslinking density.

Our measurements show that for various polymer content $\phi$, the total osmotic pressure $\Pi$ is independent of the shear modulus $G$ at low crosslink densities and decreases with $G$ at high crosslink densities (Fig. 3). According to the discussion in section 2, the $G$-independent $\Pi$ must be $\Pi_{mix}$. Then $\Pi_{ela} = \Pi - \Pi_{mix}$.

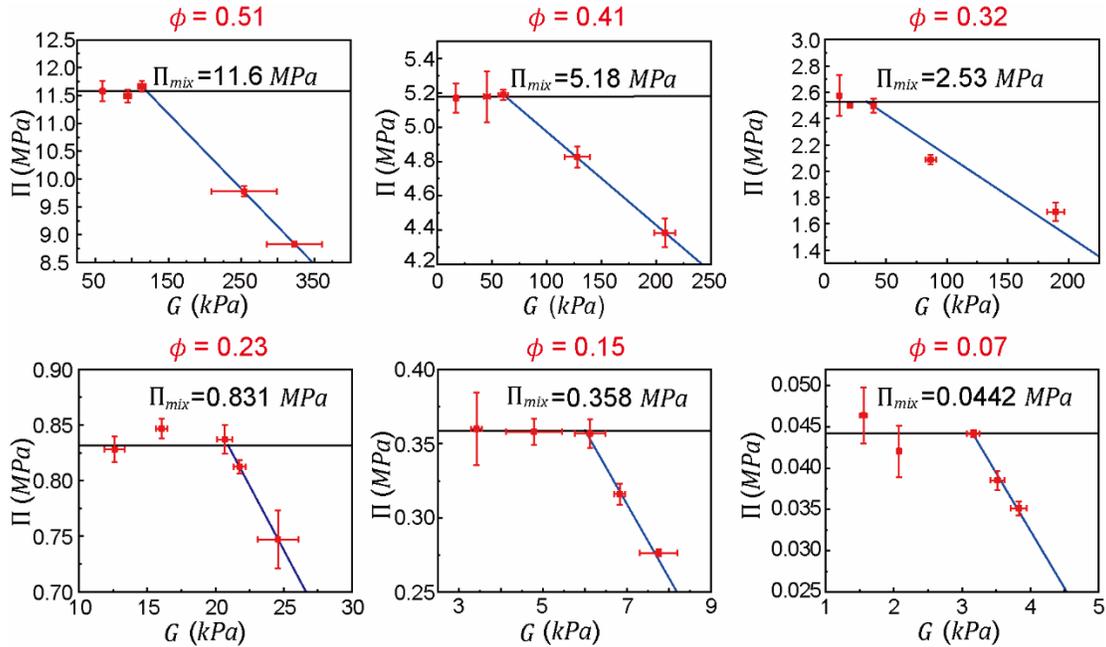

**Fig. 3 Decoupling the mixing and elastic parts in hydrogel osmotic pressure.** The relation between the osmotic pressure $\Pi$ and the shear modulus $G$ shows a plateau region and a linear region at various polymer contents $\phi$. All the error bars are for the standard error of the mean with a sample size of 3.

We compare the measured $\Pi_{mix}$ and $\Pi_{ela}$ with the Flory-Rehner model. The Flory-Rehner model predicts that $\Pi_{mix}$ is linearly proportional to $kT/\Omega$ and is a nonlinear function of $\phi$ (Eq. 7). $\Pi_{ela}$ is a function of $N$ and $\phi$ (Eq. 8), where $N$ can be converted to the shear modulus at the swollen state $G$ through $G = NkT\phi^{1/3}$ [25]. Then $\Pi_{ela}$ can be expressed as:

$$\Pi_{ela} = -G(1 - \phi^{2/3}). \tag{9}$$

Eq. 9 shows that $\Pi_{ela}$ is linearly proportional to $G$ and is a nonlinear function of $\phi$.

We first compare measured $\Pi_{ela}/G$ with Eq. 9. All quantities in Eq. 9 can be experimentally determined without fitting parameters. Our measurements show that the Flory-Rehner model significantly underestimated $-\Pi_{ela}/G$ (Fig. 4a). Moreover, the measured $\Pi_{ela}/G$ shows a non-monotonic dependence on $\phi$. This is likely because we synthesize hydrogels at various polymer contents, which results in different amounts of entanglement [56-58]. The entanglement effect has been overlooked in existing studies of $\Pi_{ela}$.

We next compare measured $\Pi_{mix}\Omega/kT$ with Eq. 7. Eq. 7 involves a fitting parameter $\chi$, which we obtain through the nonlinear least square method. We found that with a $\chi = 0.46$, the Flory-Rehner model predicts $\Pi_{mix}\Omega/kT$ reasonably well (Fig. 4b). Note that $\chi = 0.46$ is different from existing reports. In one study, Li et al. reported a series of $\Pi_{mix}$ measurements, assuming that $\Pi_{ela}$ perfectly follows the Flory-Rehner model (Fig. 4b). Compared to our results, Li's measurements under-estimated $\Pi_{mix}$. Since Li's measurements use compression tests where $\Pi_{mix}$ is balanced by elasticity at equilibrium, this under-estimation is consistent with our observation that the Flory-Rehner model underestimates $\Pi_{ela}$. Using the invalid elastic model, Li concluded that $\Pi_{mix}$ could not be fitted using the Flory-Rehner model with a constant $\chi$ even in a narrow range of polymer content $0.06 < \phi < 0.14$ (Fig. 4c). In contrast, our measurements are well-fitted by the Flory-Rehner model over a much wider range of polymer content $0.07 < \phi < 0.51$ using a constant $\chi$. Our measurements stop at $\phi = 0.51$ because the hydrogel becomes glassy at a higher $\phi = 0.62$. In another study, Day et al. reported $\chi = 0.49$ by measuring the vapor pressure of uncrosslinked pAAm solutions. However, Day's measurement only consists of five data points of extremely dilute polymer solution ($0.0014 \leq \phi \leq 0.0056$) [61]. Consequently, Day's $\chi$ value is more prone to error. In contrast, our measurement consisted of 18 measurements over a much wider range of polymer content.

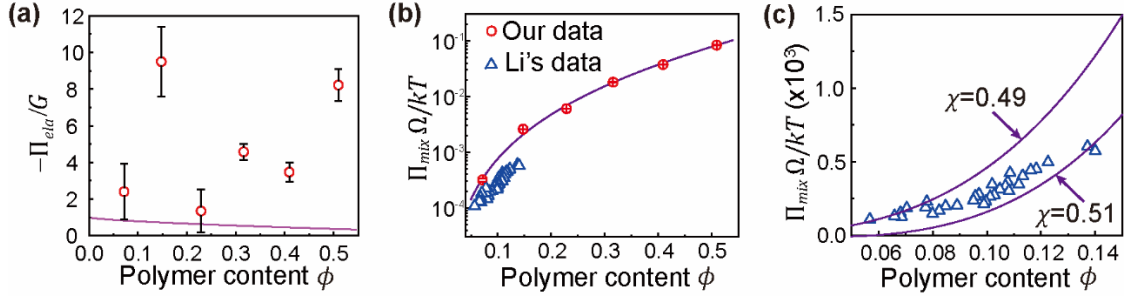

**Fig. 4 Comparing with the Flory-Rehner model. (a)** Our measurement shows the Flory-Rehner model significantly underestimates $-\Pi_{ela}/G$. **(b)** Our measurement shows $\Pi_{mix}\Omega/kT$ follows the Flory-Rehner model reasonably well with a $\chi = 0.46$. **(c)** Li's measurement, assuming the Flory-Rehner model for $\Pi_{ela}$, shows that $\Pi_{mix}\Omega/kT$ cannot be fitted with a constant $\chi$. All the error bars are for the standard error of the mean with a sample size of 3.

There have been various efforts to modify the Flory-Rehner model. For the elastic part, Treloar assumed that $W_{ela}$ follows the Neo-Hookean model without the logarithmic term in Eq. 4 [62], Then $\Pi_{ela} = -G$ instead of $\Pi_{ela} = -G(1 - \phi^{2/3})$ in Eq. 9. This modification does not change the order of magnitude of the prediction, thus not affecting the conclusion. Dai et al. have added additional fitting parameters to generate any power law relation between $\Pi_{ela}/G$ and $\phi$ [40]. However, the non-monotonic dependence in Fig. 4a cannot be described by power laws. For the mixing part, Alberto Paulin et al. suggested that $\chi$ should be an $\phi$-dependent polynomial rather than a constant.[63]. Kim et al. suggested that some strands in a polymer network do not bear loads but contribute to swelling, thus leading to an additional entropic term depending on the number of non-load-bearing chains [27]. Xu et al. suggested that entropy change during network formation will also affect the swelling behavior, thus leading to an additional entropic term depending on the polymer volume fraction at the synthesis state [28]. Our results suggest that these modifications are unnecessary and disprove the widespread belief that $\Pi_{mix}$ should be blamed for the error of the Flory-Rehner model [24-27, 63].

## 5. Conclusion

We demonstrate a novel constrained swelling test that independently characterizes the mixing and elastic parts of hydrogel osmotic pressure $\Pi$. We found that when the crosslink density is sufficiently low, the osmotic pressure is independent of elasticity, thus dominated by the mixing part, $\Pi_{mix}$. Since $\Pi_{mix}$ is independent of crosslinking, the elastic part $\Pi_{ela}$ at any crosslinking density can be determined by $\Pi - \Pi_{ela}$. The measurements show that $\Pi_{mix}$ of polyacrylamide hydrogel can be explained by the classical Flory-Rehner model reasonably well for a wide range of polymer content $0.07 \leq \phi \leq 0.51$ with $\chi = 0.46$.

On the other hand, the Flory-Rehner model predicts $\Pi_{ela}$ with order-of-magnitude error. The constrained swelling tests can be applied to other hydrogels to study more complex constitutive models. The capability to independently characterize $\Pi_{mix}$ and $\Pi_{ela}$ hints new directions in developing more accurate hydrogel constitutive models.

**Appendix**

**Synthesis of polyacrylamide hydrogel**

Acrylamide (AAm, A8887), N,N′-Methylenebisacrylamide (MBAA, M7279), and α-Ketoglutaric acid (α-Keto,75890) purchased from Sigma-Aldrich were used as a monomer, crosslinker, and photo-initiator. Deionized water was purchased from McMaster-Carr. The precursor solution is made by mixing monomer, crosslinker, and initiator in water. The monomer concentration is varied to control the polymer content $\phi$. The molar ratio between the monomer and crosslinker is varied to control the gel stiffness $G$. The hydrogel is then synthesized under UV light. Since hydrogel synthesized at high polymer content cannot attain low shear modulus due to entanglements [56], we need to further dehydrate the as-synthesized hydrogel to realize certain combinations of $\phi$ and $G$. In Table 1, the polymer contents at the final and synthesis states and the resulting shear moduli at different amounts of added crosslinkers are listed. If the polymer contents at the final and synthesis states differ from each other, the hydrogel is dehydrated before the test. Otherwise, the hydrogel is directly tested after synthesis.

**Table 1 Synthesis data for polyacrylamide hydrogels with different $G$**

| Final polymer content $\phi$ | Synthesis polymer content $\phi_0$ | Shear modulus $G$ (Molar ratios between crosslinker and monomer $n_{MBAA}:n_{AAm}$) |
|---|---|---|
| 0.51 | 0.51 | 60 kPa (0.01%), 94 kPa (0.03%), 114 kPa (0.3%), 254 kPa (2%), 323 kPa (3%) |
| 0.41 | 0.41 | 17 kPa (0.01%), 45 kPa (0.1%), 60 kPa (0.3%), 128 kPa (1%), 208 kPa (2%) |
| 0.32 | 0.07 | 12 kPa (0.6%) |
| 0.32 | 0.32 | 20 kPa (0.01%), 40 kPa (0.1%), 87 kPa (1%), 189 kPa (3%) |
| 0.23 | 0.04 | 13 kPa (0.8%) |
| 0.23 | 0.11 | 16 kPa (0.32%) |
| 0.23 | 0.23 | 21 kPa (0.01%), 22 kPa (0.02%), 23 kPa (0.04%) |
| 0.15 | 0.04 | 3.4 kPa (0.8%) |
| 0.15 | 0.07 | 4.8 kPa (0.69%) |

| | 0.15 | 0.01% (6.1), 0.02% (6.8), 0.03% (7.7) |
| --- | --- | --- |
| 0.07 | 0.04 | 1.6 kPa (0.8%), 2.1 kPa (2%) |
| | 0.07 | 3.2 kPa (0.3%), 3.5 kPa (0.5%), 3.8 kPa (0.6%) |

**Constrained swelling tests**

Stainless steel dies with 1mm depth and 5mm, 10mm, and 20mm diameters are machined from corrosion-resistant 316 stainless steel purchased from McMaster-Carr. If the as-synthesized hydrogel is directly used for the fully constrained swelling test, the hydrogel is directly synthesized in the stainless-steel die to ensure the hydrogel fully fills the volume. If the hydrogel needs to be dehydrated first, we will synthesize a hydrogel sheet with a thickness of $(\phi_0/\phi)^{-1/3}$mm, which will exactly match the 1mm thickness of the stainless-steel die after dehydration from $\phi_0$ to $\phi$. Circular hydrogel disks are then punched out from the dehydrated hydrogel sheet and fitted into the die. The diameters of the dies vary between 5mm, 10mm, and 20mm. The die with a diameter of 20 mm is paired with a 100 N load cell, and the die with diameters of 5 and 10 mm is paired with a 500 N load cell. Since osmotic pressure $\Pi$ can vary by orders of magnitude, such combination of disk size and load cell range can ensure the total force is between 10% to 100% of the load cell capacities. Stainless-steel mesh with mesh size $130\mu m \times 18\mu m$ ($200 \times 1400$ openings by 1" by 1" area) is purchased from McMaster-Carr. The mesh is washed with acetone to remove any grease. Before the test, the washed stainless-steel mesh is first placed at the bottom of the water container. The stainless-steel die with the hydrogel is glued to the load cell of a Shimadzu EZ-LX universal tester. The stainless-steel die is quickly lowered into the water bath and pressed against the metal mesh within 30s to ensure negligible swelling. A pre-compression slightly lower than the estimated osmotic pressure is applied to the metal die to ensure complete contact between all parts. After the test, the hydrogel disks are weighted. The constraint is considered sufficient if the mass change is less than 5%.